\documentstyle[11pt]{article} 

\def\APP{{\em Acta Physica Polonica} }

\def\NP{{\em Nucl. Phys.} }
\def\PLB{{\em Phys. Lett.}  B}

\def\PRD{{\em Phys. Rev.} D}
\def\ZPC{{\em Z. Phys.} C}
\def\PRC{{\em Phys. Rep.} C}
\def\IJA{{\em I.J.M.P.} A}
\def\JPN{{\em J. of Phys. G. Nucl. and Part. Phys.}}  
\def\JP{{\em J. of Physics}}
                                                                      
\def\ra{\rightarrow}

\def\be{\begin{equation}}
\def\ee{\end{equation}}
\def\bea{\begin{eqnarray}}
\def\eea{\end{eqnarray}}
\newcommand{\mr}{{\stackrel{<}{\sim}}}

\def\eg{ {\it e.g. }}
\def\ie{ {\it i.e. }}
\begin{document}
\begin{flushright}
IFT 97/20\\
DESY 97-240\\
{\large \bf hep-ph/9712518} \\
December 1997
\end{flushright}
\centerline{\large  THE STRUCTURE OF THE PHOTON IN HARD}
\centerline{{\large   HADRONIC PROCESSES}\footnote
{Lecture given at the XXXVII  School of Theoretical Physics,
Zakopane, May 30-June 10, 1997 }}
\centerline{Maria Krawczyk}
\centerline{Institute of Theoretical Physics, Warsaw University, Warsaw, 
Poland}
\centerline{and}
\centerline{Deutsches Electronen-Synchrotron DESY, Hamburg}
\begin{abstract}
{The concept of the structure of the photon is discussed
and the progress in the measurement of various
structure functions of the photon 
as well of parton distributions in the photon is shortly reviewed.}
\end{abstract}
\section{The photon and its structure}
\subsection{The notion of  the photon}
The concept of the photon originated  in 1900 in the description 
of the black body radiation when M. Planck
assumed that  the emission and the absorption
of energy should appear  in the  form of  quanta of energy. 
The Einstein suggestion presented in 1905,        
that light be considered a collection
of independent particles of energy, or particles of light,
was not easily accepted nor by Planck nor by other physicists. 
 R. A. Millikan, experimentalist working on the photo-emission from
metal surfaces,  {\sl {even in the face of his own data 
({\rm {supporting  Einstein's view}}) called it a "bold, not
 to say reckless, hypothesis"}}.
For twenty years also Bohr  resisted  the concept  of light quanta,
{\sl {he, again like Planck, 
argued that the locus of the problem was not light, but matter }}
(from [1a]). 
In 1922 a convincing evidence for light quanta appeared in 
 the  scattering of X rays on electrons (A. Compton's experiment).
The current name: the photon
was given by the American chemist G. N. Lewis in 1926.

Quantum Electrodynamics (QED), 
the theory  describing the interaction between electrons and photons,
was introduced later on (years 1925-1927 by  M. Born, W. Heisenberg, P. Jordan
and  P. Dirac [1b,c]);
the photon  plays here the role of a gauge boson, 
mediating the electromagnetic interaction.
It is assumed to be a massless and chargeless
object with a pointlike coupling to elementary, charged particles.
Its role is not changed in the Standard Model. 
No doubt, it is the oldest and the best known boson.
\subsection{The "structure" of the photon \protect \footnote{
partly based on the D. R. Yennie talk given at the XVI Zakopane School [2a]
and on [2b]}}

{\sl In quantum field theory, the electromagnetic field couples to all
particles carrying the electromagnetic current, and thus
a photon can fluctuate into virtual states of remarkable complexity.
At high energies, the fluctuation of a photon  into a Fock state of
particles of total invariant mass ${\cal {M}}$ can persist over a time of
 order
$\tau=2E_{\gamma}/{\cal M}^2$ - untill the virtual state is materialized by a
collision or annihilation with another system}, from Ref. [3g].

{\sl {At first, the photon was regarded as structureless...
As the scale of available energies increased, it was found that through 
an interaction with a Coulomb field the photon 
could materialize as pairs of electrons
\be
 \gamma \rightarrow  e^+e^-.
\ee 
Although not usually 
thought of in these terms, this phenomenon was the earliest manifestation of
photon structure.}}
So, one can say that
{\sl {the physical photon has an electron-positron pair constituent.

...The photon (real or virtual) was for purpose of hadronic 
interactions again regarded as structureless,...in reality the photon has 
an internal structure which is very similar to that of hadrons, except that 
it occurs with a probability only of order $\alpha\sim 1/137$,}} from [2b].

The hadronic properties of the photon were observed  first
 in  soft  processes like 
$\gamma p \rightarrow \rho p$ or  $\gamma p \rightarrow \gamma p$,
where  the typical for pure hadronic elastic processes falloff
with the square of the momentum transfer
$t$ was present.
Such {\sl  {soft}} hadronic
processes involving photons can be described in
the so called Vector Dominance Model(VDM), 
assuming   the "$\rho$- meson component" in the photon
(also $\omega$ and $\phi$ components, 
or  other vector mesons resonances  in the Generalized VDM (GVDM)).

As  $|t|$ increases   it is very unlikely that the process remains elastic.
The inelastic production starts to dominate, nevertheless 
one can still find in the photon-hadron 
scattering  a similarity to the  pure hadron-hadron collision. 
In both cases, for example, in the {\sl {hard}} inclusive
processes,  the quark and gluon degrees of freedom
come into the game. 
This is expected since by similar  reasoning as above, the transition
\be
 \gamma \rightarrow  q \bar {q},
\ee 
which may occur in a color field of hadronic constituents 
\footnote{it may  occur also due to a Coulomb field in  
 the process: $\gamma \gamma \ra {q \bar {q}}$ },
should be treated as a signal of the quark constituent in the photon.
The discussed above vector meson  components
of the photon arises when the $q \bar {q}$ system is  confined.  
\subsection{Parton content of the photon in QCD}
Hard hadronic processes involving partonic constituents of the photon
 can be described in Quantum Chromodynamics (QCD)
 due to smallness of the corresponding coupling
 constant $\alpha_s(Q^2)$, with  $Q^2$ being the hard scale.
 Presently such results exist  
up to  next-to-leading $\log Q^2$ terms (NLL).  

Contrary to the structure of hadrons, the structure functions for the photon 
can be calculated in the Parton Model and  already at this (Born) level  
the scaling violation appears.    The all-order
logarithmic $Q^2$ dependence 
of the partonic densities in the photon 
can  in principle be calculated in QCD in a form of the asymptotic solutions,
without the extra input at some scale, needed for hadrons. 
 A singular behaviour is obtained in the NLL calculation 
of the asymptotic solution 
at small $x_{Bj}$,  to be regularised by the 
nonperturbative  (\eg$\rho$) contribution.
  The structure function of a {\sl virtual photon}, with virtuality 
$-p^2=P^2$ in the  region where $Q^2\gg P^2\gg \Lambda_{QCD}^2$,  
is free from a such  singular behaviour at small $x_{Bj}$.
Therefore the measurement of the structure of virtual photon
 plays a  special role as a unique  test of  perturbative QCD \cite{rev}.

 {Similarly to the photon case,   one can introduce 
the partonic ``structure'' of $W/Z$ bosons \cite{szwed1} 
or leptons \cite{szwed2}.}
Note, that 
the structure of the virtual photon and the 
structure of the electron are closely related to each other
in the $e^+e^-$ or $ep$ collisions.
 This new   area
for theoretical investigations has been opened in the last few years,
leading to interesting results.

In  $e^+e^-$ collisions the dedicated DIS$_{\gamma}$ experiments
are performed in order to measure the 
 { photon } structure functions. 
Here  the photon-probe with 
the high virtuality tests the partonic
structure of the  photonic target. 
    The large $p_T$ particle or jet production in  $e^+e^-$ 
and $ep$ collisions (so called resolved photon processes)
are  suitable for this purpose as well, see \eg [3,6,7].

The existing data allow to construct the parton parametrizations for both
real and the virtual photon using  the appropriate for the 
photon 
 evolution equations (inhomogeneous ones,
 due to the direct coupling to quarks (Eq.2)).
 So far only the parametrizations for unpolarized parton densities 
are available (the review of parton parametrizations can be found in 
\cite{my}).
\subsection {The structure of the photon AD 1997}
During the last few years  a significant progress has been made 
in measurements of the structure 
function $F_2^{\gamma}$ and of the individual parton distributions in the
resolved photon processes
due to LEP and KEK $e^+e^-$ experiments, as well as 
due to photoproduction measurements at the $ep$ collider HERA
 (recent results are discussed in \eg \cite{my,stef}). 

In the single  tagged $e^+e^-$ experiments with an
 arbitrary hadronic final state
the  structure function 
$F_2^{\gamma}$ 
is measured in the $Q^2$ range between 0.24 and 390 GeV$^2$ and  
$x_{Bj}$ from 0.002 to 0.98.
Although the general behaviour of $F_2^{\gamma}$ both
 as a function of the $Q^2$
and $x_{Bj}$ agrees with the theoretical predictions,
the situation is not satisfactory.
The uncertainties of the data are   large because  of still small statistics,
and because of  
 difficulties with the unfolding of the true variables 
from 
the visible ones  (as for example  visible 
invariant mass of the hadronic system $W_{vis}$ instead of the full $W$
needed to extract  the quantity $x_{Bj}$).

Note also, that    serious discrepancies were  found recently
in the description by the existing MC generators of
some  details of the final hadronic systems
in the DIS$_{\gamma}$
experiments and 
 also in the jet production in resolved photon 
processes, both in $\gamma \gamma$ collisions  at LEP
and  in the $\gamma p$ collisions  at HERA. 
\section {Structure functions of the photon}
Following the line of reasoning from  Sec.1.2
we discuss now the structure functions of the photon.
 The cross section for the process involving
the interaction of the photon with 
elementary, charged particles can be presented symbolically 
as a series in the coupling constant
$\alpha=e^2/4\pi$:
\be
\sigma \sim \alpha+\alpha^2+\dots
\ee
For small  coupling constant,
one can approximate the cross section 
by the first, or by first few terms in the above expansion.
However  for some inclusive processes involving a large energy scale, 
 the expansion parameter may be
different - there may appear large logarithms which should then 
be summed up
to all orders.
\subsection{ Leptonic structure functions of the $\gamma$}
Let us discuss the  inclusive, pure electromagnetic process where 
the true expansion parameter is instead of 
$\alpha$ rather  $\alpha \log Q^2$ (the Leading Logarithms (LL) expansion),
and the cascade process starting  from the initial photon
(Eq.1) may be factorized (separated) from the basic hard subprocess
which occurs at a scale $Q^2$.

We will study the following  process,  where a muon pair with a 
large invariant mass is produced together with an arbitrary 
electromagnetic state $X$: 
\be 
\gamma e^+ \rightarrow \mu^+ \mu^- X ~(\rm ~leptons~ and ~photons).  
\ee
The leading order (LO) cross section for  process (4) is given by: 
\be
\sigma_{\gamma e^+\rightarrow \mu^+ \mu^- X}(s,M^2)=\int dx_{\gamma}
 f_{e/\gamma}(x_{\gamma},Q^2)
 {\hat {\sigma}}  
_{e^+e^-\rightarrow  \mu^+ \mu^-}(M^2).
\ee
The function $f_{e/\gamma}(x_{\gamma},Q^2)$
describes  the probability (within the LL accuracy in the LO approach) 
to find in the initial photon an electron 
with a fraction of  momentum $x_{\gamma}$, at the scale $Q^2$.
${\hat {\sigma}}$ is here the lowest order  cross section for  the  
  muon pair  production ($\sim \alpha^2$)  with large 
invariant mass $M^2$, which  serves here as the scale for the 
large logarithms, $Q^2$=$M^2$.

The electromagnetic structure functions of the photon related 
to the introduced above function
$f$ are being  measured presently in the following
Deep Inelastic Scattering on photon (DIS$_{\gamma}$):
\be
e(k) \gamma(p) \rightarrow e(k') X ~(\rm ~~leptons),
\ee
at a scale $Q^2=-q^2=-(k-k^{\prime})^2$,  usually greater that 1 $
~{\rm GeV}^2$.
\footnote{the  limit 1 GeV$^2$ arises since
 the discussed measurement 
is in practice correlated to the one, where the QCD structure functions
is probed (see below for details)}
 
In the single tagged events at  $e^+e^-$ colliders 
the initial (target) photon is almost
real, \ie $P^2=-p^2 \ll 1 {~\rm GeV}^2 $ (see Fig.1). 
\begin{figure}[ht]      
\vskip -0.3in\relax\noindent\hskip 1.05in
       \relax{\includegraphics{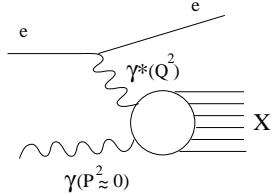}}
\vspace{19ex}
\baselineskip 0.4cm
\caption{ {Deep Inelastic Scattering on a real photon, 
$p^2=-P^2\approx 0$.}}\label{fig:dis}
\end{figure}  
To describe the DIS$_{\gamma}$ process (6)   
the following  variables are being used:
\be
 x_{Bj}={{Q^2}\over {2p\cdot q}} , 
 \hspace{2cm} y={{p\cdot q} \over {p\cdot k}}.  
\ee
(Note that in the LO approach  $x_{Bj}=x_{\gamma}$).
The differential cross section   for process (6), for unpolarized
initial particles,  is given by the following 
 QED or leptonic structure functions:
\be
{{d\sigma}\over{dx_{Bj}dy}} = {{4\pi\alpha^2}\over {Q^4}}2p\cdot k
[(1-y)F_2^{\gamma(QED)}(x_{Bj},Q^2)+x_{Bj}y^2 F_1^{\gamma(QED)}(x_{Bj},Q^2)].
\ee
Note that  the function  $F_1
^{\gamma(QED)}$(equal to the transverse $F_T^{\gamma(QED)}$) 
or the longitudinal function $F_L^{\gamma(QED)}$ 
 ($F_L^{\gamma(QED)}=F_2^{\gamma(QED)}-2x_{Bj}F_T^{\gamma(QED)}$)
 are not easily accessible,
due to the small $y$ range probed in present experiments.

Some of the recent data for the $F_2^{\gamma(QED)}$, obtained  for the 
muonic final state, are presented in Fig. 2
together with the QED prediction, based on the first order
process:
$$ \gamma^* \gamma \rightarrow \mu^+ \mu^{-}. $$
(Other  structure functions (azimuthal correlations), which arise when
final state particles are observed  were measured as well, see discussion in
 \cite{my,stef}.)
\vspace*{2cm}
\begin{figure}[ht]      
  \vskip -0.2in\relax\noindent\hskip -1.705in
       \relax{\includegraphics{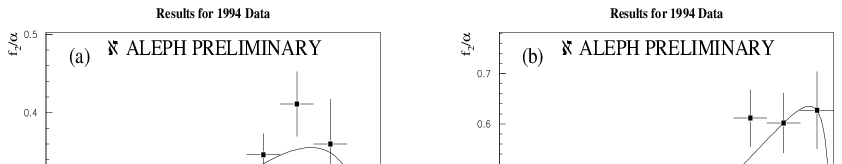}} 
       \relax\noindent\hskip 3.505in
       \relax{\includegraphics{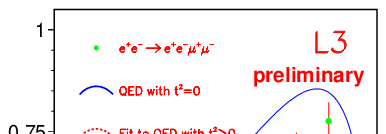}}
\vspace{22ex}
\baselineskip 0.4cm
\caption{ {The leptonic structure function of the  photon for the different 
$P^2$ values, a) and b) ALEPH data at $Q^2$=2.790 and 14.649 GeV$^2$ 
[8a]; c) L3 data at $Q^2$=3.25 GeV$^2$ [8b] }} 
\end{figure}

The important additional results from these 
measurements is the estimation of the 
averaged virtuality of the initial photon needed for the extraction of
hadronic structure functions, see below.
\subsection{ Hadronic (QCD) structure functions of the $\gamma$}
Let us assume now that in the  first step  the photon
decays  with  probability
$\alpha$ into a pair of quark  antiquark (Eq.2).
Then the subsequent radiation processes will be rather
governed by the strong coupling constant $\alpha_s$
than by the electromagnetic one. For the
inclusive production of hadrons   
the true expansion parameter is expected to be
  $\alpha_s \log Q^2$,
with the $Q^2$  scale parameter  being,
in order to apply the perturbative QCD, 
   larger than $\Lambda_{QCD}^2$.
Then the cascade process originated from the initial photon
 can be described in the perturbative QCD in terms of the 
 parton distribution in the photon.
 The analogue of the process (4) may be now the process:
\be
 \gamma {\bar {q}} \ra \mu^+ {\mu}^{-} X {(\rm {hadrons})},
\ee
with the LO formula
 for the cross section
\be
\sigma_{\gamma {\bar q}\rightarrow \mu^+ \mu^- X}(s,M^2)
=\int dx_{\gamma}f_{q/\gamma} (x_{\gamma},Q^2)
 {\hat {\sigma}}  
_{q {\bar q}\rightarrow  \mu^+ \mu^-}(M^2),
\ee
where the function $f_{q/\gamma}(x_{\gamma},Q^2)$
describes  the probability within the LL accuracy 
to find in the initial photon a quark 
with a fraction of  momentum $x_{\gamma}$, at the scale $Q^2$.
The hard process here is the Drell-Yan
process for  muon pair  production with large 
invariant mass $M^2$,
and $Q^2$=$M^2$. (See the Secs. 3.1 and 3.2,
 where other  hard processes
"resolving"  the photon are discussed.)

When in the final state only hadrons are produced
in the DIS$_{\gamma}$ experiment at $e^+e^-$ colliders, 
the (hadronic) structure functions
of the photon  $F^{\gamma}_{1,2..}$ 
related to $f_{q/\gamma}$ are measured.
Since only part of the final hadronic state is observed
in practice, the proper estimation of $P^2$ 
and also the proper unfolding of the true variables,
 \eg $x_{Bj}=Q^2/(Q^2+W^2+P^2)$,
is crucial. 

Below we discuss separately the case of a real (or almost real)
photon (with $P^2\mr \Lambda_{QCD}^2$) and the case of a virtual photon,
 where $Q^2 \gg P^2 \gg \Lambda_{QCD}^2$.
\subsubsection{Real photon}
The unpolarized deep inelastic scattering on the real  photon,
\be
e(k) ~~\gamma(p) \ra e(k^{\prime}) ~~X(\rm hadrons), \label{eq:dis}
\ee
with a large momentum transfer between the electrons: 
$Q^2=-q^2=-(k-k^{\prime})^2\gg 1 ~{\rm GeV}^2$, can be described by two
independent (hadronic) structure functions 
$F^{\gamma}_{1}$ ${\rm ~and~}$
$~F^{\gamma}_{2} {\rm~or}$ $~F^{\gamma}_{L}$, according to 
 Eq.8.
 The following formula 
which relates the structure function to the quark densities 
holds in LO approach 
(here $x_{\gamma}=x_{Bj})$:
\be
{{F_2^{\gamma}(x_{Bj},Q^2)}\over{x_{Bj}}}=
\sum_q^{2N_f}e_q^2 f_{q/\gamma} (x_{Bj},Q^2)
={{\alpha}\over{2\pi}}N_c\sum_q^{2N_f}e_q^4 [x^2_{Bj}+(1-x^2_{Bj})]\log 
{{Q^2}\over{\Lambda^2_{QCD}}}. 
\ee
The  existing results for $F_2^{\gamma}$ as a function of
   $x_{Bj}$ (from \cite{stef})  and of   $Q^2$
 are  shown in Figs 3a) and 3b), respectively.  
Note, that the low $x_{Bj}$ behaviour of $F_2^{\gamma}$ 
still has to be clarified,
as parton parametrizations give different predictions here.
\newpage
\vspace*{15cm}
\begin{figure}[ht]      
\vskip 0.in\relax\noindent\hskip -1.75in
       \relax{\includegraphics{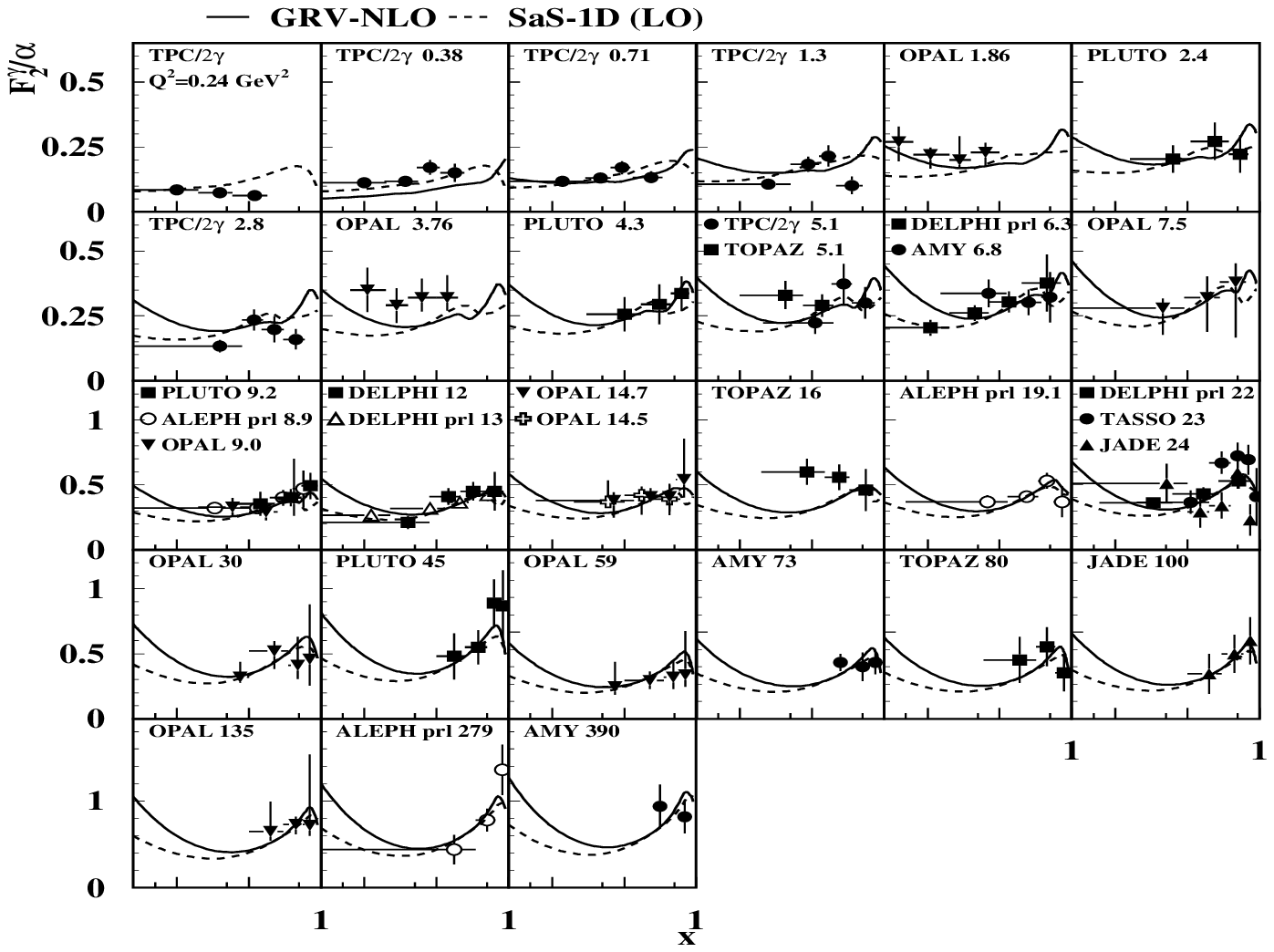}}
       \relax\noindent\hskip 3.3in
       \relax{\includegraphics{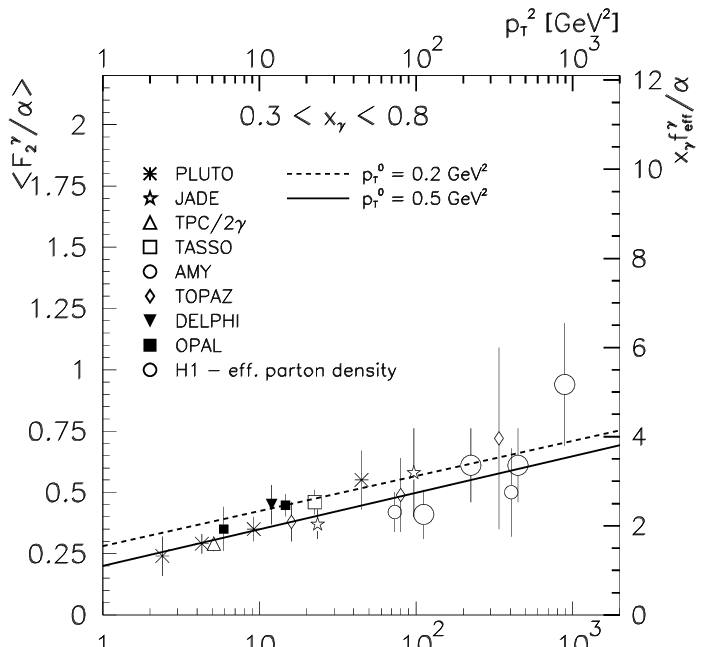}}
\vspace{0.ex}
\baselineskip 0.4cm
\caption{{ {a) The $x_{Bj}$ 
dependence of $F_2^{\gamma}$ with the
 predictions of the GRV-NLO and Sas-1D(LO) parton parametrizations (from [7]);
b) the $Q^2$ dependence of  $F_2^{\gamma}$ averaged on the $x_{Bj}$
 range between 0.3 and 0.8,
 together with data from HERA (H1), 
based on the effective parton density, from [9]. }}} 
\label{fig: New data }
\end{figure}
\subsubsection{Virtual photon}
In the region where $Q^2 \gg P^2 \gg \Lambda_{QCD}^2$, 
the structure of the virtual photon may be tested. 
The Parton Model (PM) formula for the corresponding structure
 function $F_2^{\gamma}$
contains a $ \log  Q^2/P^2$ term (instead $\log{{Q^2}/{\Lambda_{QCD}^2}}$,
 see Eq. 12), and will disappear when both scales
 approach each other. The higher order QCD corrections will not change 
 this behaviour.
There are no new data on the structure function of the virtual photon.
Fig.4 shows  the only existing  (PLUTO) data and the comparison with the 
PM, VDM and QCD predictions.  

\begin{figure}[ht]      
\vskip 2.7in\relax\noindent\hskip -0.705in
       \relax{\includegraphics{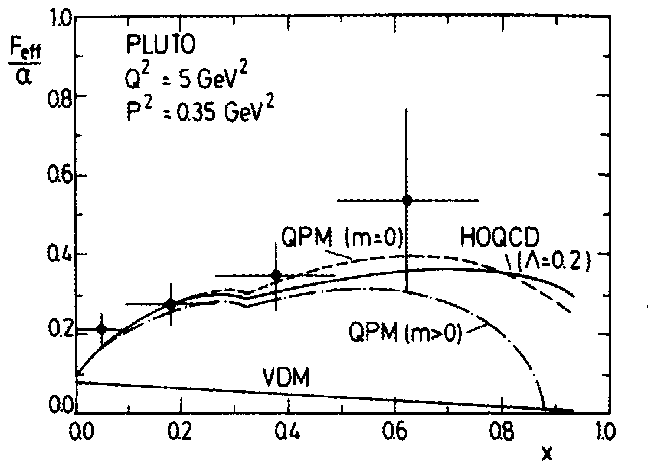}}
\relax\noindent\hskip 2.505in
       \relax{\includegraphics{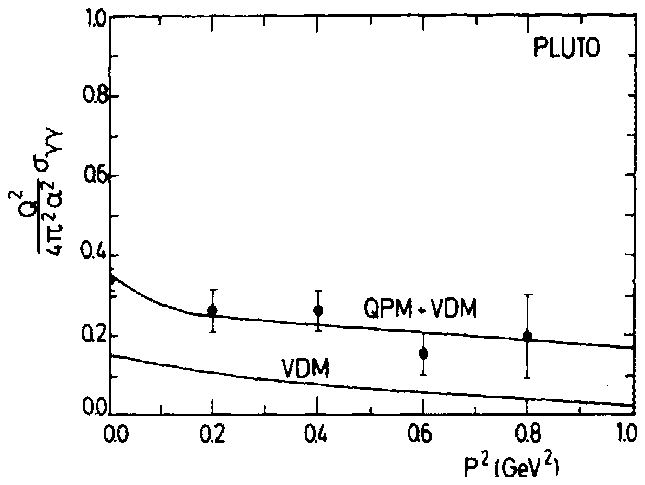}}
\vspace{-12ex}
\baselineskip 0.4cm
\caption{ {a) the $x_{Bj}$ dependence, 
b) the dependence of the $F_2^{\gamma}$ on  $P^2$  for the virtual photon 
 averaged $Q^2$ and $x_{Bj}$ ranges, from [10] }} 
\end{figure}  
\section{Resolved photon processes}
 Large $p_T$
 particles production can be used to measure the 
partonic content of the photon. For a  discussion on the newest results
see \eg \cite{stef}. Below we discuss  the 
most important resolved photon processes, namely those involving jets
(see also \cite{my}).
\subsection
{Jet production with large $p_T$}
\begin{figure}[ht]      
\vskip 3.45in\relax\noindent\hskip .35in
       \relax{\includegraphics{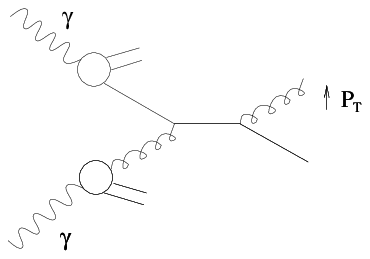}}
             \relax\noindent\hskip 1.95in
        \relax{\includegraphics{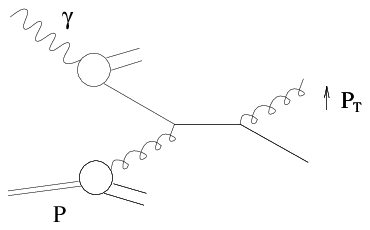}}
\vspace{-36ex}
\baselineskip 0.4cm
\caption{ {Resolved photon processes in a) $\gamma \gamma$
and b) $\gamma p$ collisions. }} 
\end{figure}
Measurements of the production of jets with large transverse momentum
 in the (resolved) real or virtual photon processes 
give a complementary to the DIS$_{\gamma}$ experiments
 information on the parton density in the photon, 
being \eg much more sensitive to the gluon density.
Such analyses
are performed now in 
$e^+e^-$ experiments as well as at the $ep$ HERA collider.
In case of the $\gamma \gamma$ processes direct photon
 (\ie without the partonic "agent"), single and double resolved 
photon processes are studied, whereas in the $\gamma p$ case 
only direct and single resolved ones.
In Figs. 5a and 5b examples of resolved photon processes in $\gamma \gamma$
and $\gamma p$ collisions are presented.
 The relevant $x_{\gamma}$ distributions
of the initial photons, with  $x_{\gamma}\sim 1$ expected for the direct
contribution,  are shown  in Figs.6a and 6b.
\begin{figure}[ht]      
\vskip 1.5in\relax\noindent\hskip -1.25in
        \relax{\includegraphics{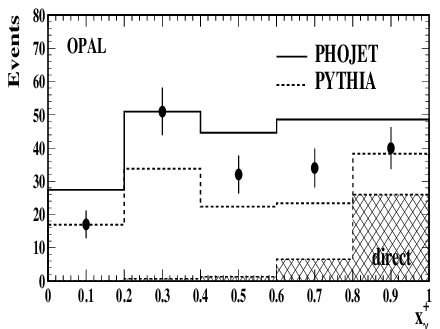}}
\relax\noindent\hskip 1.25in
          \relax{\includegraphics{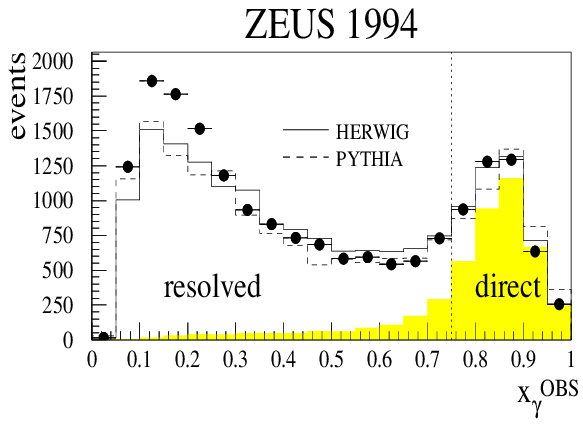}}
              \relax\noindent\hskip 2.25in
              \relax{\includegraphics{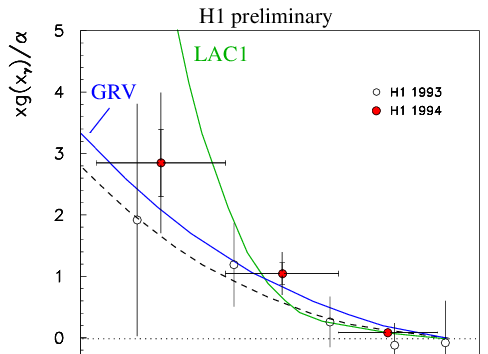}}
\vspace{3ex}
\baselineskip 0.4cm
\caption{ {The $x_{\gamma}$ distribution from a) OPAL [11a], b) ZEUS [11b];
c) the gluon density at $Q^2$ =75 GeV$^2$ from H1 [9] compared
 to the LAC1 and the GRV-LO parametrizations.}} 
\end{figure} 
The gluon distribution in the real photon 
extracted from the jet production  data for
$Q^2$ =$p_T^2$=75 GeV$^2$ at HERA is shown in Fig.6c.
The effective parton densities were also measured at HERA, the constructed
from them  the effective structure function 
$F_2^{\gamma}$ is plotted in Fig.3b.
\begin{figure}[ht] 
\vskip 7.cm               \relax\noindent\hskip 2.25in
              \relax{\includegraphics{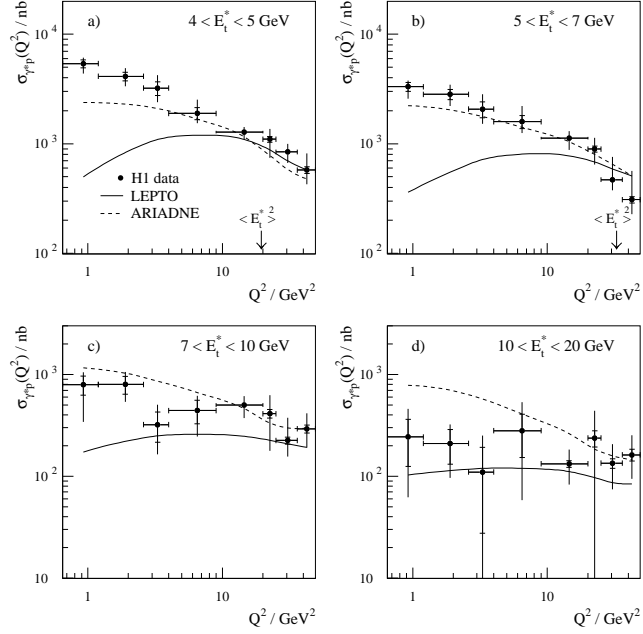}}
\vspace{7ex}
\baselineskip 0.4cm
\caption{ {The dependence of the cross section
$\sigma_{\gamma^*p}$ on the squared 
virtuality of the initial photon as  measured at HERA (H1), from [12]. 
 }} 
\end{figure} 

In the resolved photon processes the content of the virtual photon 
can be studied as well. The cross section measurement 
for the different virtualities  of the photon was performed at the 
HERA collider.
The hard $Q^2$ scale corresponds here to the transverse energy of jets, 
$E_T^2$.
Only if $E_T^2$   is bigger than the virtuality 
squared for the initial photon
the interpretation in terms of the structure function 
(parton distributions) of the virtual photon is appropriate 
(see Fig.7 for results, to be compared with Fig.4b).
\subsection{Compton scattering $\gamma p \ra \gamma X$}
\begin{figure}[ht] 
\vskip 3.5cm               \relax\noindent\hskip 2.05in
          \relax{\includegraphics{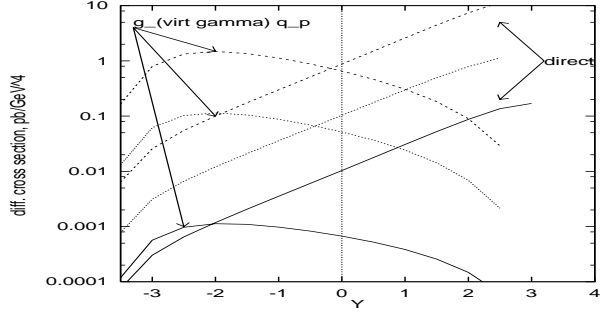}}
\vspace{5.5ex}
\baselineskip 0.4cm
\caption{ {The rapidity  distribution 
in the $\gamma p$ centr of mass system at HERA for  the photon produced 
with $p_T$=5 GeV for initial photon virtualities: 
$P^2$=0.03, 0.25 and 2.5 GeV$^2$(upper, middle and lower lines)  
 [13c].}} 
\end{figure} 
The large $p_T$
 photon produced in the Deep Inelastic Compton (DIC) process 
may be used to study the content of the photon as well \cite{pro}.
Note that recently this process, with almost real initial photon,
was measured at HERA [14].  
 In papers [13b,c] we study the possibility of probing the structure
 of the virtual photon in DIC scattering at HERA.
 Fig. 8 shows the domination of the process 
$g_{\gamma^*}q_p \ra \gamma q$ over the direct contribution:
${\gamma}^*q_p \ra \gamma q$, for  different virtualities of the initial
photon. This result suggests a possibility
 to measure the gluonic content of the virtual photon in DIC process at HERA 
[13c].
\section{Summary}
 An  impressive  progress was made in the last few years in the measurements
of the structure functions and individual parton distributions in the photon,
both in $e^+e^-$ and $ep$ experiments.
Still more data are  needed in order to clarify 
 the small $x_{Bj}$ behaviour of $F_2^{\gamma}$, to measure 
 the polarized parton densities,
and to test the structure of the virtual photon. 
 The interplay between the structure of the electron and of the 
virtual photon may also be important in future analyses.\\ 
 Being an  important  test of QCD, the structure of the photon may  be also  
 a useful tool in the high energy physics
in studying  the effects of  
 "new physics", as  due to the partonic content of photon
a new production mechanisms   may appear.\\ 
{\bf Acknowledgments}\\
I wish to thank the organizers of this excellent School and of all previous 
Zakopane Schools I was happy to attend.
I am indebted to Peter Zerwas for the important comment on the early 
developments of QED and pointing me the reference [1b].
I am also very grateful to Aaron Levy for 
a critical reading of the manuscript.
 I wish to  thank 
Andrzej Zembrzuski for his help in preparing 
this contribution and  
 Stefan S\"{o}lndner-Rembold  for sending his newest compilation on
 $F^{\gamma}_2$ data.
Supported in part by the Polish Committee for Scientific Research, 
Grant No 2P03B18410.


\baselineskip 0.4cm
\begin{thebibliography}{99}
\bibitem{light}a) A. Zajonc, Catching the light , Oxford University Press, 
Oxford New York, 1995, Chapters 9 and 10;
b) P. Jordan, talk at Neutrino Conference, Aachen 1976, in proc. p. 494;
c) S. S. Schweber, QED and the Men Who Made It:
 Dyson, Feynman, Schwinger, and Tomonaga, 
Princeton University Press, Princeton, 1994.
\bibitem{yenni}a) D. R. Yenni, \APP {\bf B7}, 897 (1976), 
 b) T. H. Bauer et al., Rev.Mod.Phys. 
{\bf 50}, 261 (1978); Erratum ibid {\bf 51}, 407 (1979).
\bibitem{rev}
a) V. M. Budnev et al. , \PRC {\bf 15}, 181 (1975);
b)C. Peterson, T. F. Walsh, P. M. Zerwas, \NP B{\bf 174}, 424 (1980);
c)H. Kolanoski , {\em Springer Tracts in Modern Physics}` 84 - 85; 
d)Ch. Berger and W. Wagner,  \PRC 146 (1987) 1;
e)H. Abramowicz et al. , \IJA {\bf8}, 1005 (1993);
f)M. Drees and R. M. Godbole , Pramana - \JP ~~{\bf 41}, 83 (1993);
M. Drees and R. M. Godbole , \PRD {\bf 50}, 3124 (1994); \JPN {\bf 21}, 1559
 (1995);
g)S. Brodsky and P. Zerwas, {\sl Nucl. Instr. and Methods in Phys.Res}.
{\bf A355}, 19 (1995);
\bibitem{szwed1} W. S\l omi\'{n}ski and J. Szwed, \ZPC {\bf 72},87 (1996), 
\PRD {\bf 52},1650(1995), \PLB {\bf 323},427(1994).
\bibitem{szwed2}W. S\l omi\'{n}ski and J. Szwed, \APP B{\bf27}, 1887(1996), 
\PLB {\bf 387},861(1996); M. Drees and R. Godbole, \PRD {\bf 50},3124 (1994). 
\bibitem{my} M. Krawczyk, M. Staszel and A. Zembrzuski,
 Survey on the photon structure 
functions and the resolved photon processes, preprint IFT 15/97, July 1997.
\bibitem{stef} S. S\"{o}lndner-Rembold, 
Talk at XVIII International Symposium on Lepton-Photon Interaction,
Hamburg, Germany, July 28-August 1, 1997.
\bibitem{f2qed}a) ALEPH Coll., C. A. Brew, S. Cartwright, M. Lehto,
 Proc. of PHOTON'97, Egmond aan Zee, The Netherlands, May, 1997;
 b) L3 Coll., submitted to EPS'97, Jerusalem, August 1997.
\bibitem{Miller} K. M\"{u}ller, talk at EPS'97, Jerusalem, August 1997,
based on the H1 results, T. Ahmed at al., \NP B{\bf 445}, 195 (1997)
  and paper 270 submitted to   EPS'97.
\bibitem{f2wirt} PLUTO Coll., Ch. Berger et al., \PLB {\bf 142}, 119 (1984).
\bibitem{xgamma}a) OPAL Coll., K. Ackerstaff et al., \ZPC {\bf 73}, 433
 (1997);
 b) ZEUS Coll., J. M. Butterworth et al., preprint 97-04 UCL/HEP,
in Proc. of the Ringberg Workshop, Germany, May 1997.
\bibitem{h1virt} H1 Coll.,C. Adloff at al., DESY 97-197 (hep-ex/9709017).
\bibitem{pro}a) M. Krawczyk, \APP B{\bf21}, 999 (1990); A. Bawa, M. Krawczyk,
 W.J. Stirling, \ZPC
{\bf 50}, 293 (1991);
P. Aurenche at al., HERA Workshop, Hamburg 1987, p.561
 and \ZPC {\bf 56}, 589 (1992); 
L. E. Gordon and J. K. Storrow, \ZPC {\bf 63}, 581 (1994); b)
   M. Krawczyk, A. Zembrzuski, Physics at HERA Workshop, Hamburg 
1991, vol.1, p. 617;
c) M. Krawczyk, A. Zembrzuski \PRD {\bf 57} (1998)
\bibitem{dic} ZEUS Coll., paper 656 submitted to EPS'97, Jerusalem, 
August 1997.
\end{thebibliography}
\end{document}